\documentclass[10pt,journal,compsoc]{IEEEtran}

\usepackage{multirow}
\usepackage{tikz}
\usetikzlibrary{tikzmark}
\usepackage{todonotes}
\usepackage{algorithm}
\usepackage{algpseudocode}
\usepackage[utf8]{inputenc}
\usepackage{tikz}
\usepackage{pgf-umlsd}
\usepackage{subfig}
\usepackage{mathrsfs}

\usepackage{amssymb}% http://ctan.org/pkg/amssymb
\usepackage{pifont}% http://ctan.org/pkg/pifont
\newcommand{\cmark}{\ding{51}}%
\newcommand{\xmark}{\ding{55}}%

\ifCLASSOPTIONcompsoc
  \usepackage[nocompress]{cite}
\else
  \usepackage{cite}
\fi

\begin{document}

\title{\textit{Delays have Dangerous Ends}: Slow HTTP/2 DoS attacks into the Wild and their Real-Time Detection using Event Sequence Analysis}

\author{Nikhil Tripathi, ~\IEEEmembership{Member,~IEEE}%,  Abhijith Kalayil Shaji
        % John~Doe,~\IEEEmembership{Fellow,~OSA,}
        % and~Jane~Doe,~\IEEEmembership{Life~Fellow,~IEEE}% <-this % stops a space
\IEEEcompsocitemizethanks{\IEEEcompsocthanksitem N. Tripathi is with the Computer Science \& Engineering Group at Indian Institute of Information Technology, Sri City, Chittoor, India 517646.\protect\\
E-mail: nikhil.t@iiits.in
%\IEEEcompsocthanksitem A. K. Shaji is with the facutly of Computer Science at Otto von Guericke University, 39106 Magdeburg, Germany. \protect\\E-mail: abhijith.kalayil@yahoo.com
}% <-this % stops a space
% \thanks{Manuscript received April 19, 2005; revised August 26, 2015.}
}

\IEEEtitleabstractindextext{%
\begin{abstract}
The robustness principle, written by Jon Postel in an early version of TCP implementation, states that the communicating entities should be \textit{liberal} while accepting the data. Several entities on the Internet do follow this principle. For instance, in this work, we show that many popular web servers on the Internet are generous as they wait for a substantial time period to receive the remaining portion of an incomplete web request. Unfortunately, this behavior also makes them vulnerable to a class of cyber attacks, commonly known as Slow Rate DoS attacks. HTTP/2, the recent version of HTTP, is recently found vulnerable to these attacks. However, the impact of Slow HTTP/2 DoS attacks on real web servers on the Internet has not been studied yet. Also, to the best of our knowledge, there is no defense scheme known to detect Slow Rate DoS attacks against HTTP/2 in real-time. To bridge these gaps, we first test the behavior of HTTP/2 supporting web servers on the Internet against Slow HTTP/2 DoS attacks. Subsequently, we propose a scheme to detect these attacks in real-time. We show that the proposed detection scheme can detect attacks in real-time with high accuracy and marginal computational overhead.

\end{abstract}

\begin{IEEEkeywords}
HTTP/2, Slow Rate DoS attacks, Anomaly detection, DoS
\end{IEEEkeywords}}

\maketitle

\IEEEdisplaynontitleabstractindextext
\IEEEpeerreviewmaketitle

\ifCLASSOPTIONcompsoc
\IEEEraisesectionheading{\section{Introduction}\label{sec:introduction}}
\else
\section{Introduction}
\label{sec:introduction}
\fi

\IEEEPARstart{A}{pplication} layer Denial-of-Service (DoS) attacks are a newer class of DoS attacks that exploit vulnerabilities present in the application layer protocols such as HTTP, NTP, DHCP \cite{tripathi2016secure, tripathi2016probabilistic, tripathi2021preventing}. These attacks target specific services (e.g., web, mail, file transfer) running on a victim computer with a relatively slighter impact on the network resources \cite{Mantas, gonzalez2014impact}. One famous example of application layer DoS attacks is the Slow Rate DoS attack. Slow Rate DoS attack requires sending multiple incomplete requests to the victim server to consume its connection queue space. Once this finite space is consumed, the server can not entertain further incoming requests. A few instances of Slow Rate DoS attacks have also been encountered in the past \cite{imperva_slowloris, tripathi2021application}. 

HTTP/1.1 is one of the most studied protocols against Slow Rate DoS attacks \cite{tripathi2021application, tripathi2016secure}. HTTP/2, the successor of HTTP/1.1, was launched in May 2015 \cite{rfc7540}, and our HTTP/2 usage measurement study (see Section \ref{sec:http2_usage}) shows that approximately 71\% of the websites serve web content using HTTP/2. The operational differences between HTTP/1.1 and HTTP/2 make the Slow Rate DoS attacks against HTTP/1.1 \cite{tripathi2016secure} ineffective against HTTP/2. Likewise, the defense mechanisms known to counter Slow Rate DoS attacks against HTTP/1.1 are ineffective in detecting Slow Rate DoS attacks against HTTP/2. In one of our prior works \cite{tripathi2018slow}, we performed the vulnerability assessment of HTTP/2, and proposed five new Slow Rate DoS attacks that are effective against it. However, we tested the attacks in a controlled lab setup only, and we did not observe the behavior of real HTTP/2 web servers against them. Thus, it it is essential to evaluate the effects of Slow HTTP/2 DoS attacks on the real web servers deployed on the Internet. Taking motivation from it, in this work, we test the behavior of HTTP/2 supporting web servers on the Internet against Slow HTTP/2 DoS attacks, and showed that several popular web servers are vulnerable to them\footnote{The empirical evaluation results are also presented in our recent preliminary work \cite{tripathi2022}.}. We also propose a detection mechanism in this work to proactively detect these attacks. We make the following specific contributions in this paper:
\begin{itemize}
    \item We perform a comprehensive Internet-wide study by scanning the top 500K Alexa websites to find how many of them support HTTP/2. Our study shows that approximately 71\% of the top 500K websites support HTTP/2. 
    \item We test the behavior of HTTP/2 supporting web servers on the Internet against different Slow HTTP/2 DoS attacks and show that several of them are vulnerable to the attacks.
    \item We propose an event sequence analysis-based detection scheme to detect Slow HTTP/2 DoS attacks in real-time.
    \item We test the detection performance of the proposed scheme in a real network and show that it can detect attacks with very high accuracy and marginal computational overhead.
\end{itemize}

The rest of the paper is organized as follows. We present the HTTP/2 overview and the literature review in Sections \ref{sec:prelim} and \ref{sec:literature}. In Sections \ref{sec:http2_usage} and \ref{sec:evaluation}, we then discuss our study to measure HTTP/2 usage on the Internet and the empirical evaluation of Slow HTTP/2 DoS attacks, respectively. In Section \ref{sec:detection}, we discuss the proposed detection scheme and present the experiments conducted to test the detection performance of the proposed scheme in Section \ref{sec:exp_eval}. Finally, the paper is concluded in Section \ref{sec:conclusion}.

 

\section{Preliminaries}
\label{sec:prelim}

This section presents some background knowledge on HTTP/2 and Slow Rate DoS attacks.

\subsection{HTTP/2}
\label{subsec:http2}

HTTP/2 is defined in RFC 7540 \cite{rfc7540} and recently became a standard in May 2015. HTTP/2 operation differs significantly from its predecessor HTTP/1.1 \cite{8090408, 8090411}. HTTP/2 was proposed keeping in mind the inability of HTTP/1.1 to utilize the TCP’s transmission capacity efficiently. HTTP/2 achieves higher efficiency by using several virtual connections known as \textit{streams} over a single TCP connection. HTTP/2 also mitigates head-of-line blocking of the web requests that persists in HTTP/1.1. HTTP/2 also implements its own flow control mechanism at the application layer. It is required to prevent the communicating entities from overwhelming each other by sending data simultaneously through multiple streams over a single TCP connection. HTTP/2 also involves using different types of frames for different purposes \cite{rfc7540}. For example, HTTP/2 HEADERS and CONTINUATION frames are used to transmit the header part of the web request, while HTTP/2 DATA frames are used to transmit the message body (e.g., HTML code of a web page). HTTP/2 SETTINGS frames are used to negotiate connection parameters, and HTTP/2 GOAWAY frames are used to close the connection. HTTP/2 WINDOW\_UPDATE frame indicates the size of data (in bytes) that the sender can accept from the receiver in addition to the current window size. These frames are carried inside the TCP payload and should not be confused with the link-layer frames. Interested readers are requested to refer to RFC 7540 \cite{rfc7540} for the list of HTTP/2 frames and their usage.

\subsection{Slow Rate DoS Attacks}
\label{subsec:slow_dos}

Slow Rate DoS attacks \cite{zargar2013survey} require sending multiple specially crafted incomplete requests to a victim server. On receiving such requests, the server stores them in its connection queue space and waits for a time duration $\mathcal{T}$ (predefined in the configuration file) in the hope of receiving the remaining portions of the web request. However, the attacker never sends the remaining portion of the complete requests. To consume the server's connection queue space, the attacker establishes enough connections, and from each connection, it sends incomplete requests. Once the connection queue space is consumed, the server can not entertain further incoming requests, resulting in DoS scenario. It is not a good approach to mitigate these attacks by terminating the TCP connections with a connection duration greater than a predefined threshold. This is because incomplete requests are not uncommon on the Internet \cite{tripathi2016secure, tripathi2018slow}, and such an approach may also terminate the connections from the benign clients with poor TCP connection quality. Examples of such clients are remote clients with high latency and clients on low-grade cellular or satellite networks. 
%These attacks are highly stealth and thus, are difficult to detect as compared to other DoS/DDoS attacks against HTTP/2 \cite{10.1145/3407023.3409198, tripathi2018slow, tripathi2021application}.

\section{Related Works}
\label{sec:literature}

Several works in the literature discuss DoS/DDoS attacks against HTTP/1.1 and the defense mechanisms to counter those attacks \cite{tripathi2021application, zargar2013survey, praseed}. However, the same does not hold true for its successor, HTTP/2. Since HTTP/2 is a recently standardized protocol, only a few works discuss possible attacks against HTTP/2 and the defense mechanisms to counter them \cite{tripathi2021application}. Moreover, since HTTP/2 operation differs significantly from HTTP/1.1 \cite{8090408, 8090411}, the existing techniques to detect Slow Rate DoS attacks against HTTP/1.1 \cite{tripathi2016secure, core, antiloris, limitipconn, reqtimeout} are ineffective in detecting Slow Rate DoS attacks against HTTP/2. As a result, researchers in the security community recently proposed a few defense approaches to detect Slow HTTP/2 DoS attacks. This section first mentions DoS/DDoS attacks known against HTTP/2 and, subsequently, the defense mechanisms to counter them.

\subsection{DoS/DDoS attacks against HTTP/2}
\label{subsec:dosddos}

Imperva \cite{imperva_http2} disclosed some implementation vulnerabilities in HTTP/2, which can be exploited for creating situations such as Blue Screen of Death and arbitrary code execution at the client side. However, these vulnerabilities existed in the protocol implementations but not in the protocol itself. Post the vulnerability exposure; the affected implementations were patched to mitigate these vulnerabilities. Beckett and Sezer \cite{8090408, 8090411} studied how HTTP/2 functionalities can be exploited to launch flood and amplification-based DDoS attacks. Adi et al. \cite{Adi2016} tested how the HTTP/2 web servers behave if control frame flooding is launched against them. Praseed and Thilagam \cite{9171310, praseed2019multiplexed} demonstrated that the request multiplexing feature of HTTP/2 can be exploited to launch asymmetric DDoS attacks on the web servers. In particular, the attacker sends high workload web requests resulting in heavy computational overhead at the web server. The multiplexed asymmetric attack may result in either partial DoS, also known as Reduction-of-Quality (RoQ), or a complete DoS scenario. The attacks discussed in these works are distributed in fashion and, thus, typically fall under the Distributed Denial-of-Service (DDoS) category \cite{tripathi2021application}. Since these attacks require sending a large number of web requests to the server to render it unusable, they are relatively difficult to launch and easier to detect \cite{10.1145/3407023.3409198}. %Moreover, these attacks can be detected using techniques that monitor parameters such as a sudden surge in the amount of incoming traffic at the server, abnormal computational resource consumption, etc. Different variations of such detection approaches are discussed comprehensively in some of the recent review papers \cite{praseed, zargar2013survey, singh2017application}. 

Unlike traditional DDoS attacks, Slow HTTP/2 DoS attacks \cite{tripathi2018slow} need small computational power for their execution at the attacker’s side. They involve sending very few requests to the web server to prevent it from serving benign clients \cite{10.1145/3407023.3409198, tripathi2021application}. Since these attacks generate a smaller amount of traffic, they are stealthier and, thus, difficult to detect as compared to traditional DDoS attacks \cite{10.1145/3407023.3409198, zargar2013survey, tripathi2021application}. \textbf{Taking motivation from this, we restrict our focus only to Slow HTTP/2 DoS attacks in this paper.} 

Based on the HTTP/2 frame parameters that are altered, there can be different variants of Slow HTTP/2 DoS attacks. Interested readers are requested to refer to our previous work \cite{tripathi2018slow} for a detailed working of these attacks. For completeness, we present a summary of these attacks below:

\subsubsection{Advertising Zero Window Size (Attack-1)}

In this attack, the attacker sends an HTTP/2 SETTINGS frame to the server to inform that it does not have any window space available to receive the data. The server does not send any data back to the attacker on receiving this frame unless it receives an HTTP/2 WINDOW\_UPDATE frame from the attacker. However, the attacker never sends the required WINDOW\_UPDATE frame to the server forcing the latter to wait.

\subsubsection{Incomplete POST Request Message Body (Attack-2)}

HTTP/2 uses DATA frames to carry the message body of a web request (e.g., values in an HTML form). A DATA frame defines an END\_STREAM flag. This flag, when \textit{set}, indicates that the frame carries the entire message body and will not be followed by any other DATA frame. However, to launch this attack, the attacker sends a DATA frame having END\_STREAM flag \textit{reset}. On receiving this frame, the server waits to receive more DATA frames such that the last DATA frame must have the END\_STREAM flag set.
However, the attacker never sends the subsequent DATA frames, forcing the web server to wait for the complete message body.

\subsubsection{Sending Connection Preface Only (Attack-3)} 

Connection preface is used to inform a web server that HTTP/2 will be used for further communication. The attacker only sends the connection preface to launch this attack and does not send any subsequent HTTP/2 frames. On receiving the connection preface only, the server waits to receive further HTTP/2 frames belonging to the web request.

\subsubsection{Incomplete GET/POST Request Header (Attack-4)}

HTTP/2 HEADERS frames carry the header part of the web requests. If the header is large enough not to fit in a single HEADERS frame, it is split into smaller parts such that the first header part is sent into the HEADERS frame, while the remaining parts are sent into the CONTINUATION frames. A HEADERS frame defines an END\_HEADERS flag. This flag, when \textit{set}, indicates that the HEADERS frame carries the entire header, and any CONTINUATION frame will not follow it. However, to launch this attack, the attacker sends a HEADERS frame having the END\_HEADERS flag \textit{reset}. On receiving this frame, the server waits to receive CONTINUATION frames such that the last CONTINUATION frame must have the END\_HEADERS flag set. However, the attacker never sends the subsequent CONTINUATION frames, forcing the web server to wait for the complete header.

\subsubsection{Unacknowledged SETTINGS frame (Attack-5)}

On receiving an HTTP/2 SETTINGS frame, a recipient (client/server) must acknowledge it by sending an empty SETTINGS frame back to the sender. The attacker does not acknowledge an HTTP/2 SETTINGS frame sent by the server to launch this attack. As a result, the web server keeps waiting to receive the SETTINGS frame acknowledgment.

\subsection{Known Defense Mechanisms}

Several works in the literature discuss defense strategies to counter attacks against HTTP/1.1. However, those strategies can not detect attacks against HTTP/2 due to the several operational differences between HTTP/1.1 and HTTP/2 \cite{8090408, 8090411}. Moreover, since HTTP/2 is a newly standardized protocol, only a few existing works focused on HTTP/2 security. In one of our previous works \cite{tripathi2018slow}, we proposed an offline anomaly detection scheme to detect if a web server received incomplete web requests in a particular time interval. However, the detection scheme could detect if the interval contains attack traffic only after it is elapsed. Thus, it could not detect the attacks in real-time. In another work, Praseed and Thilagam \cite{praseed2021fuzzy} proposed an approach to model an HTTP/2 request set as a fuzzy multiset and subsequently assign a trust score to distinguish between valid and invalid request sets. This approach could detect the multiplexed asymmetric DDoS attack with decent accuracy. However, it was not designed to detect Slow HTTP/2 DoS attacks.

%In\cite{muraleedharan2021deep}, authors proposed a deep neural network classification technique to detect Slow Rate DoS attacks. The authors used the CICIDS 2017 dataset \cite{cicids} for testing the detection performance of the scheme and showed that it could detect the attacks with a very high accuracy. However, the proposed technique was restricted to HTTP/1.1 only and it could not detect Slow Rate DoS attacks against HTTP/2.

\subsection{Empirical Evaluation of Slow HTTP/2 DoS Attacks}

In our recent preliminary work \cite{tripathi2022}, we tested the behavior of HTTP/2 supporting web servers on the Internet against Slow HTTP/2 DoS attacks. We showed that several popular web servers are vulnerable to these attacks. However, we did not present any defense approach in that work which can be deployed to counter Slow HTTP/2 DoS attacks.

Table \ref{table:literature_comp} briefly summarizes how our work differs from the prior works related to HTTP/2 security.

\begin{table}
\renewcommand{\arraystretch}{1.3}
\caption{Comparison with prior works}
\label{table:literature_comp}
\centering
\begin{tabular}{|p{3.5cm}|p{2.3cm}|p{1.7cm}|}
\hline
References & Empirical Evaluation of Slow HTTP/2 DoS & Detection of Slow HTTP/2 DoS \\ \hline
Adi et al. \cite{Adi2016} & \xmark & \xmark \\ \hline
Tripathi and Hubballi \cite{tripathi2018slow} & \xmark & \cmark (only offline) \\ \hline
Praseed and Thilagam \cite{9171310} & \xmark & \xmark \\ \hline 
Praseed and Thilagam \cite{praseed2019multiplexed} & \xmark & \xmark \\ \hline
Beckett and Sezer \cite{8090408} & \xmark & \xmark \\ \hline
Beckett and Sezer \cite{8090411} & \xmark & \xmark \\ \hline
Praseed and Thilagam \cite{praseed2021fuzzy} & \xmark & \xmark \\ \hline
Tripathi and Shaji \cite{tripathi2022} & \cmark & \xmark \\ \hline
\textbf{This paper} & \cmark & \cmark \\ \hline
\end{tabular}
\end{table}

\section{Measuring HTTP/2 Usage on the Internet}
\label{sec:http2_usage}

In September 2021, we conducted a comprehensive study to measure HTTP/2 usage on the Internet. We took into account the top 500,000 Alexa domains in this study and tested if they support HTTP/2. For this purpose, we wrote a python script that communicates with the web server of each domain in the Alexa list. In particular, the python script sends a ‘Client Hello’ message to the web server as part of the standard TLS handshake. Through this message, the script informs the web server about the application protocols that it supports. Examples of these application protocols are HTTP/2 over TLS (known as h2), HTTP/2 over cleartext (known as h2c), and HTTP/1.1. In response, the server also sends a `Server Hello' message to the script, thereby informing about the application protocols that it supports.

We conducted this measurement study into four iterations. In the first iteration, we collected HTTP/2 usage statistics by performing TLS handshakes with the web servers of the top 500K domains in the Alexa list. In this iteration, we received responses from 445315 web servers, out of which 335854 web servers were found to support HTTP/2. Moreover, 109461 web servers were found to support HTTP/1.1. The remaining 54685 web servers did not respond in the first iteration for two reasons. First, the TCP connections were timed out as there was no response from some web servers. Second, the authoritative name servers for some of the domains have an \texttt{A} record for the hostname \texttt{www} only. For example, the authoritative name servers for a domain, say, \texttt{iiits.ac.in} have the record \texttt{www.iiits.ac.in A 103.21.58.130}, but they do not have the record \texttt{iiits.ac.in A 103.21.58.130}. Thus, the web server \texttt{103.21.58.130} is reachable only when a web request is sent to the Fully Qualified Domain Name (FQDN) \texttt{www.iiits.ac.in}. We also observed that many domains support HTTP/1.1, but they serve web content over HTTP/2 if a user visits the FQDN \texttt{www.<domain>} as the web server mapped to the FQDN \texttt{www.<domain>} supports HTTP/2, while the web server mapped to \texttt{<domain>} supports HTTP/1.1 only. The result of the first iteration measurement is shown in Table \ref{table:http2_usage}.

\begin{table*}[h]
\renewcommand{\arraystretch}{1.3}
\caption{HTTP usage statistics. h2: HTTP/2 over TLS, h2c: HTTP/2 over cleartext, http/1.1: HTTP/1.1, NA: Not Applicable}
\label{table:http2_usage}
\centering
\begin{tabular}{|p{1.3cm}|p{1cm}|p{3cm}|p{3.75cm}|p{2.75cm}|p{3.5cm}|}
\hline
\multicolumn{2}{|c|}{} & 1st iteration (\texttt{https://$<$domain$>$})& 2nd iteration (\texttt{https://www.$<$domain$>$})& 3rd iteration (\texttt{http://$<$domain$>$}) & 4th iteration (\texttt{http://www.$<$domain$>$})\\ \hline
\multirow{3}{*}{Responded} & h2 & \textbf{335854} & \textbf{017220} & NA & NA \\ \cline{2-6}
& h2c & NA & NA & \textbf{002791} & \textbf{000552} \\ \cline{2-6}
& http/1.1 & 109461 \tikzmark{a} & 101841 \tikzmark{d} & 000000 & 000000\\ \hline
\multicolumn{2}{|c|}{Not responded} & 054685 \tikzmark{b} & 045085 \tikzmark{e} & 144135 \tikzmark{g} & 143583 \\ \hline
\multicolumn{2}{|c|}{Total} & 500000 & \tikzmark{c} 164146 & \tikzmark{f} 146926 & \tikzmark{h} 144135\\ \hline
\end{tabular}
\begin{tikzpicture}[overlay, remember picture, shorten >=.5pt, shorten <=.5pt, transform canvas={yshift=.25\baselineskip}]
    \draw [->] ({pic cs:a}) to ({pic cs:c});
    \draw [->] ([yshift=.75pt]{pic cs:b}) -- ({pic cs:c});
    \draw [->] ({pic cs:d}) to ({pic cs:f});
    \draw [->] ([yshift=.75pt]{pic cs:e}) -- ({pic cs:f});
    \draw [->] ({pic cs:g}) to ({pic cs:h});
  \end{tikzpicture}
\end{table*}

In the second iteration, we appended the \texttt{www} keyword to all those domains (=164146) whose web servers either supported HTTP/1.1 (=109461) or did not respond (=54685) in the first iteration. This provided us FQDNs of the format \texttt{www.<domain>}. Subsequently, we collected HTTP/2 usage statistics by performing TLS handshakes with the web servers corresponding to these FQDNs. The result of the second iteration measurement is shown in Table \ref{table:http2_usage}. It can be noticed from the table that 17220 more HTTP/2 supporting domains were found in the second iteration. Also, the number of domains that either supported HTTP/1.1 (=101841) or were unresponsive (=45085) was decreased in the second iteration.

The popular web clients such as Chrome, Firefox, and Safari support h2 only; however, the debugging and benchmarking tools such as curl and h2load support h2 as well as h2c. Thus, popular server-side implementations such as Apache simultaneously support both h2c and h2 configurations to avoid compatibility issues. In the first and second iterations, we measured the usage of h2 on the Internet using the \texttt{HTTPS} scheme. In the third iteration, we measured the usage of h2c on the Internet using the \texttt{HTTP} scheme. We checked the HTTP/2 support only for those domains which either supported HTTP/1.1 or did not respond in the second iteration. To measure the h2c usage statistics, our script sends a cleartext HTTP/2 web request to a web server after the 3-way TCP handshake and subsequently checks if the response from the web server is an HTTP/2 response. The result of the third iteration measurement is shown in Table \ref{table:http2_usage}. It can be noticed from the table that no response was received for 144135 domains. Thus, in the fourth iteration, we appended the \texttt{www} keyword to the unresponsive domains in the third iteration and obtained FQDNs of the format \texttt{www.<domain>}. Subsequently, we collected HTTP/2 statistics by sending cleartext HTTP/2 web requests to the web servers corresponding to these FQDNs. The result of the fourth iteration measurement is shown in Table \ref{table:http2_usage}. It can be noticed from the table that the web servers corresponding to \textbf{356417} (sum of the numbers shown in bold fonts in Table \ref{table:http2_usage}) domains were found to support HTTP/2.

\section{Empirical Evaluation of the Attacks}
\label{sec:evaluation}

\begin{figure}
\centering
\includegraphics[width=3.5in]{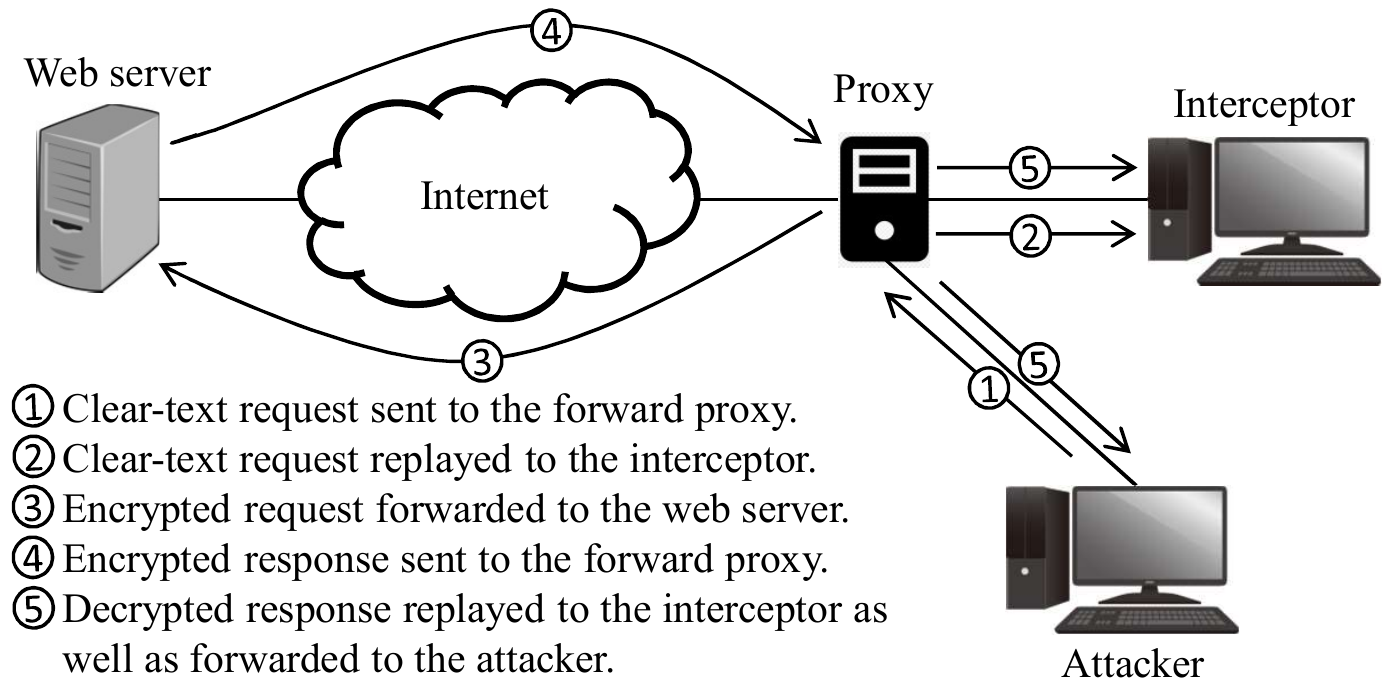}
\caption{Testbed setup}
\label{testbed}
\end{figure}

This section presents an empirical evaluation of Slow HTTP/2 DoS attacks. We created a testbed setup for this evaluation, as shown in Figure \ref{testbed}. 

\subsection{Testbed Setup}

Our testbed had four entities - 1) an attacker machine, 2) a transparent forward proxy, 3) an interceptor, and 4) the target web server. 
\\
\textbf{Attacker:} We designated a computer as the attacker machine and executed on it the attack scripts written in Python. These attack scripts sent the web servers exactly one incomplete HTTP/2 request of each attack (Attack-1 to 5) type. We sent \textit{only} one incomplete request to observe the servers' behavior, and we did not intend to launch the attacks against the servers by sending multiple attack requests.
\\
\textbf{Transparent Forward Proxy:} As discussed in the previous section, most of the web servers on the Internet support HTTP/2 over TLS (h2). Thus, to intercept and monitor otherwise encrypted HTTPS traffic, we configured a transparent forward proxy \cite{netresec} in front of the attacker machine such that the proxy connects to the web servers on behalf of the attacker. The proxy takes cleartext HTTP/2 requests from the attacker as input and establishes TLS connections\footnote{The TLS connections are established only with h2 servers. In the case of h2c servers, the proxy simply forwarded the cleartext requests.} with the web server of the domain present in the \texttt{HOST} field of the HTTP/2 requests. Subsequently, it forwards the encrypted web request to the server. On receiving the encrypted HTTP/2 response from the server, the proxy decrypts the response and forwards it to the attacker. It also replays the cleartext HTTP/2 traffic from the web server and the attacker to the interceptor, as shown in Figure \ref{testbed}. Before replaying the traffic to the interceptor, the proxy modifies the IP addresses appropriately such that the ultimate endpoints of the traffic become the web server and the attacker. The proxy uses the \texttt{PCAP-OVER-IP} method to replay the traffic to the interceptor on a designated port 57012. In this way, we intercepted the cleartext HTTP/2 traffic so that it could be analyzed later.
\\
\textbf{Interceptor:} We designated a computer as the interceptor that intercepted the cleartext HTTP/2 traffic replayed by the proxy. The interceptor machine captured the inbound traffic to port 57012 using tcpdump and stored it in a \texttt{pcap} file. 
\\
\textbf{Target Web Server:} The target web server belongs to the 356417 HTTP/2 supporting web servers that we found on the Internet. %We tested the behavior of each of these web servers against Slow Rate DoS attacks. On receiving the attack specific web requests from the proxy, the web servers respond back to it. 

\subsection{HTTP/2 Traffic Analysis}

\begin{figure}
\centering
\includegraphics[width=3.75in]{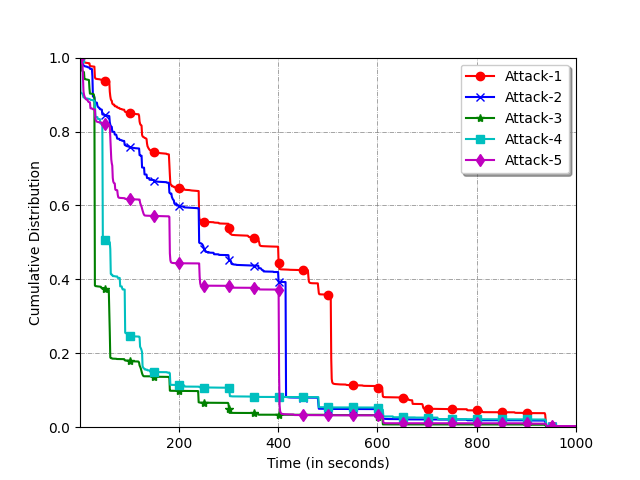}
\caption{CDF plot of the waiting time of top 500K Alexa websites. Please refer to the color version for better illustration.}
\label{cdf_plot}
\end{figure}

The captured HTTP/2 traffic at the interceptor was analyzed to compute the connection waiting time for different target web servers. For each flow\footnote{We differentiated the TCP flows using four parameters - source IP, source port, destination IP, and destination port.} in the \texttt{pcap} file, we calculated the time difference between the connection establishment (3-way handshake) and its termination (FIN-ACK packet exchange). In this way, we obtained the time the web server waited before closing the connection from which an incomplete request was sent. A CDF plot of the duration for which different web servers wait in case of each attack is shown in Figure \ref{cdf_plot}. 
We can notice from the figure that approximately 3\% of the web servers waited for more than 360 seconds before closing a connection from which an attack-3 type web request was sent. However, approximately 50\% of the web servers were found to wait for the same duration before closing the connection from which an attack-1 type web request was sent. Since attack-1 can make more web servers wait for a particular time duration, it is more effective than attack-3. The other similar inferences can also be drawn from the graph shown in Figure \ref{cdf_plot}.

\section{Proposed Detection Scheme}
\label{sec:detection}

An HTTP/2 request involves exchanging different types of HTTP/2 frames, as discussed in Section \ref{subsec:http2}. If we treat the exchange of these frames as unique events that occur one after the other, we can obtain an event sequence that is possible during the interaction between a client and an HTTP/2 server. Since the frames exchanged during the HTTP/2 request depend upon various parameters such as the user's browsing behavior and available window size at the receiver, several distinct event sequences are possible. All possible normal event sequences can be stored into a database of characteristic normal patterns (observed sequences of exchanged HTTP/2 frames) and subsequently be used to detect anomalous sequences. An HTTP/2 request is considered as attack request if the event sequence generated corresponding to it does not exist in the normal database and vice versa. However, simply checking the non-existence of an event sequence in the database to detect attack requests may result in several false positives. This is because there can be many possible unique event sequences (see Section \ref{subsubsec:database_size}), and maintaining a database of all the sequences to capture the complete normal behavior of HTTP/2 would be a tedious task.

To avoid this issue, we maintain a database of possible lookahead pairs extracted from the possible event sequences. Later in Section\ref{subsubsec:database_size}, we show that the database of possible \textbf{lookahead pairs} can capture the normal behavior of HTTP/2 with more completeness than the database of possible \textbf{event sequences}. Once we obtain the normal database of lookahead pairs during the learning phase, we assign a mismatch score to an event sequence during the detection phase based on the number of lookahead pairs extracted from it that do not exist in the normal database. The event sequence is declared anomalous (and thus, the request from which it is generated, is declared as attack request) only when its mismatch score exceeds a predefined threshold value. 
%This approach provides to our detection strategy more granular control over the classification of the sequences as normal or anomalous. 
%In the following subsections, we describe the working of our proposed detection scheme in detail.

\subsection{Learning Phase}
\label{subsec:learning}

During the learning phase, our detection scheme builds a normal database of lookahead pairs extracted from the possible event sequences and a database of maximum normal delay possible between the occurrences of two consecutive events. Algorithm \ref{algo:eve_seq} describes the procedure for generating these databases during the learning phase.
\begin{algorithm}
\caption{Learning Phase}
\label{algo:eve_seq}
    \hspace*{\algorithmicindent} \textbf{Input:} $\mathcal{F}$ - Clear-text HTTP/2 flows; $n$ - Window size \\
    \hspace*{\algorithmicindent} \textbf{Output:} $\mathcal{D}_{lookahead}$, $\mathcal{D}_{delay}$
    \begin{algorithmic}[1]
        \State $\mathcal{D}_{lookahead}$=\{\}; $\mathcal{D}_{delay}$=\{\}; $i$=1;
        \For{$flow$ in $\mathcal{F}$}
            \State $seq$=`$Start\rightarrow *$'; %$prev_{event}$=`$Start\rightarrow *$';
            \For{$frame$ in $flow$}
                \State $e$=$frame.translateToEvent()$;
                \State $seq$+=`$\rightarrow e  \rightarrow*$';
                %\State $prev_{event}$=$event$;
            \EndFor
            \State $seq$+=`$\rightarrow End$';
            \State $l$=$seq.extractLookaheadPairs()$
            \State $\mathcal{D}_{lookahead}.append(l)$
            \State $events$ = $seq.extractEvents()$;
            \While{$i <= len(events)-1$}
                \State $delay$=$events[i].time()$-$events[i-1].time()$;
                \State $str$=`$events$[$i$-1] $\rightarrow$ $events$[$i$]';
                \If{$str$ not in $\mathcal{D}_{delay}$}
                    $\mathcal{D}_{delay}$[str]=$delay$;
                \Else
                    \State $\mathcal{D}_{delay}$[str]=$max$($delay$, $\mathcal{D}_{delay}$[str]);
                \EndIf
                \State $i$+=1;
            \EndWhile
        \EndFor
        \State return $(\mathcal{D}_{lookahead}, \mathcal{D}_{delay})$;
    \end{algorithmic}
\end{algorithm}
Below we discuss the procedure for generating these databases in detail.

\subsubsection{Bulding the lookahead pairs database ($\mathcal{D}_{lookahead}$)}
\label{subsec:lookahead_extraction}

During the learning phase, we first collect normal HTTP/2 traffic and subsequently construct individual TCP flows by combining TCP payloads of all the packets exchanged after the connection establishment and before the connection termination. Once flows are constructed, they are parsed using Deep Packet Inspection (DPI) to spot HTTP/2 frames. These frames are then translated into events, as shown in Table \ref{table:frames_to_events}.
%We rely on HTTP/2 traffic collected from a real web server on the Internet rather than formal specification of the protocol's expected behavior. The benefit of this approach is that we need not determine the normal behavior of HTTP/2 from the server software implementation. Instead, we can simply capture it by collecting normal traces from an HTTP/2 web server. 
%\footnote{It is to be noted that the header and the payload of HTTP/2 frames constitute the TCP payload.} 
\begin{table}[h]
\renewcommand{\arraystretch}{1.3}
\caption{HTTP/2 frames and translated events}
\label{table:frames_to_events}
\centering
\begin{tabular}{|p{5cm}|p{3cm}|}
\hline
Frame & Event \\ \hline
3-way handshake & $Start$ \\ \hline
Connection preface & $Pref$ \\ \hline
DATA frame with End Stream (ES) reset & $Data\_frame\_!ES$ \\ \hline
DATA frame with ES set & $Data\_frame\_ES$ \\ \hline
HEADERS frame with ES and End Header (EH) reset & $Hdr\_frame\_!(ESEH)$ \\ \hline
HEADERS frame with ES reset and EH set & $Hdr\_frame\_!ES\_EH$ \\ \hline
HEADERS frame with ES and End Header (EH) set & $Hdr\_frame\_(ESEH)$ \\ \hline
%RST STREAM frame & $Rst\_stream$ \\ \hline
%PING frame & $Ping$ \\ \hline
%PRIORITY frame & $Priority$ \\ \hline
Acknowledgement to SETTINGS frame & $Settings\_ACK$ \\ \hline
SETTINGS frame with non-zero length & $Settings\_UNACK$ \\ \hline
SETTINGS frame having initial window size=0 & $Ini\_Win\_Size=0$ \\ \hline
SETTINGS frame having initial window size!0 & $Ini\_Win\_Size!0$ \\ \hline
SETTINGS frame having maximum concurrent stream!0 & $Max\_Con\_Strm!0$ \\ \hline
SETTINGS frame having maximum concurrent stream=0 & $Max\_Con\_Strm=0$ \\ \hline
WINDOW frame with increment in window size!0 & $win\_size\_incr!0$ \\ \hline
WINDOW frame with increment in window size=0 & $win\_size\_incr=0$ \\ \hline
GOAWAY frame & $GOAWAY$ \\ \hline
CONTINUATION frame & $CONTINUATION$ \\ \hline
Connection termination & $End$  \\ \hline
\end{tabular}
\end{table}
%We particularly take into account only those HTTP/2 frames that are essential to i) capture the normal behavior of HTTP/2 and ii) differentiate the benign requests from the Slow HTTP/2 DoS attack requests.
After translating the HTTP/2 frames into the corresponding events, we append the events one after the another to generate an event sequence. For the event sequence generation, we consider a TCP connection's 3-way handshake as the beginning of the flow (represented by event \textit{Start}) and the connection termination (represented by event \textit{End}) as the end of the flow. One sample event sequence for a TCP flow \texttt{IP$_A$:port$_A$} $\leftrightarrow$ \texttt{IP$_B$:port$_B$} carrying an HTTP/2 request is as follows:
\begin{quote}
   $Start \rightarrow * \rightarrow Pref \rightarrow * \rightarrow Max\_Con\_Strm!0 \rightarrow Ini\_Win\_Size!0 \rightarrow * \rightarrow win\_size\_incr!0$ 
\end{quote}
In this sequence, `$*$' event corresponds to the end of an HTTP2 frame, while '$\rightarrow$' is not an event and only represents a transition from one event to another.

Once an event sequence is generated, the lookahead pairs are extracted from it and stored in $\mathcal{D}_{lookahead}$. For this, we slide a window of size $n+1$ across the event sequence and capture which events follow which within the sliding window. For example, if we consider $n = 3$ and the following event sequence:

\begin{quote}
    $Start \rightarrow * \rightarrow Max\_Con\_Strm!0 \rightarrow Ini\_Win\_Size!0 \rightarrow * \rightarrow win\_size\_incr!0 \rightarrow * \rightarrow Pref$
\end{quote}

For each event, we capture the event that comes after it at spot 1, spot 2, and so on, up to spot $n$, as we slide the window across the above event sequence. For the first window $w_1$, from index 1 (\textit{*}) to index 3 (\textit{Ini\_Win\_Size!0}) in the sequence, we obtain the database having six lookahead pairs as shown in Table \ref{table:lookahead_w1}.
\begin{table}
\renewcommand{\arraystretch}{1.3}
\caption{Lookahead pairs for $w_1$}
\label{table:lookahead_w1}
\centering
\begin{tabular}{|p{1.75cm}|p{1.75cm}|p{1.75cm}|p{1.75cm}|}
\hline
& Index-1 & -2 & -3 \\ \hline
$Start$ & $*$ & $Max\_Con\_$ $Strm!0$ & $Ini\_Win\_Size$ $!0$ \\ \hline
$*$ & $Max\_Con\_$ $Strm!0$ & $Ini\_Win\_Size$ $!0$ & \\ \hline
$Max\_Con\_$ $Strm!0$ & $Ini\_Win\_Size$ $!0$ &  &  \\ \hline
\end{tabular}
\end{table}

Similarly, after sliding the window across the complete event sequence, we obtain the lookahead pairs as shown in Table \ref{table:lookahead_all}. 
\begin{table}
\renewcommand{\arraystretch}{1.3}
\caption{Lookahead pairs extracted from the complete event sequence}
\label{table:lookahead_all}
\centering
\begin{tabular}{|p{1.75cm}|p{1.75cm}|p{1.75cm}|p{1.75cm}|}
\hline
& Index-1 & -2 & -3 \\ \hline
$Start$ & $*$ & $Max\_Con\_$ $Strm!0$ & $Ini\_Win\_Size$ $!0$ \\ \hline
$*$ & $Max\_Con\_$ $Strm!0$ \textbf{;} $win\_size\_incr$ $!0$ \textbf{;} $Pref$ & $Ini\_Win\_Size$ $!0$ \textbf{;} $*$ & $*$ \textbf{;} $Pref$ \\ \hline
$Max\_Con\_$ $Strm!0$ & $Ini\_Win\_Size$ $!0$ & $*$ & $win\_size\_incr$ $!0$ \\ \hline
$Ini\_Win\_Size$ $!0$ & $*$ & $win\_size\_incr$ $!0$ & $*$ \\ \hline
%$*$ & $win\_size\_incr$ $!0$ & $*$ & $Pref$ \\ \hline
\end{tabular}
\end{table}
A lookahead pair is stored in the database in the form of `$event[i]:event[j],k$' where $k \leq n$ (window size) and corresponds to the distance between the $i^{th}$ and $j^{th}$ events in a sequence. Once the database of lookahead pairs ($\mathcal{D}_{lookahead}$) is built, our detection scheme refers to it to assign anomaly scores to the event sequences generated from incoming HTTP/2 requests during detection phase. 

\subsubsection{Bulding the maximum delay database ($\mathcal{D}_{delay}$)}
\label{subsec:avg_delay}

The proposed detection scheme also maintains a database, $\mathcal{D}_{delay}$, of the maximum normal delay possible between the occurrences of two consecutive events, $\mathcal{A}$ and $\mathcal{B}$. This is essential to detect the determined attackers who send the benign HTTP/2 frames to the server but with a prolonged delay between them to launch the Slow Rate DoS attacks. An entry in the maximum delay database is stored in the form `$event[i] \rightarrow event[i+1]=d$' where $d$ is the maximum delay between the $i^{th}$ and $(i+1)^{th}$ events in a sequence. %In the next subsection, we discuss how the proposed scheme detects the anomalous sequences during the detection phase.

\subsection{Detection Phase}

Algorithm \ref{algo:detection} describes the procedure for detecting an anomalous event sequence during the detection phase. We use different threads for executing the functions responsible for sniffing the incoming packets and extracting events from them, checking timeouts between the events, and assigning anomaly scores to the sequences. However, for brevity, we omit these details in Algorithm \ref{algo:detection}. The proposed scheme generates an event sequence $seq$ for each incoming HTTP/2 request during the detection phase using the method discussed in Section \ref{subsec:lookahead_extraction}. During this phase, the scheme also inserts a `timeout string' after an event $\mathscr{e}$ in $seq$ if a time duration $t$ passes after the occurrence of $\mathscr{e}$ such that $t$ is greater than the maximum delay possible between the occurrence of $\mathscr{e}$ and any other event that can occur immediately after $\mathscr{e}$. The maximum possible delay value is retrieved from $\mathcal{D}_{delay}$ database generated during the learning phase, as discussed in Section \ref{subsec:avg_delay}. %This timeout checking is described in the \texttt{checkTimeout} procedure of Algorithm \ref{algo:detection}. 

\begin{algorithm}
\caption{Detection Phase}
\label{algo:detection}
    \hspace*{\algorithmicindent} \textbf{Input:} $\mathcal{D}_{lookahead}$; $\mathcal{D}_{delay}$; $n$ - Window size; $\mathscr{t}$ - Mismatch score maximum threshold\\
    \hspace*{\algorithmicindent} \textbf{Input:} $\mathcal{F}$ - Incoming TCP flow\\
    \hspace*{\algorithmicindent} \textbf{Output:} $D$ - Declare sequence as anomalous or normal
    \begin{algorithmic}[1]
        \State $seq$ = `';
        \While{True}
            \State Continuously sniff packets belonging to $\mathcal{F}$;
            \For{$frame$ in $\mathcal{F}$}
                \State $e$=$frame.translateToEvent()$;
                \State $seq$+=`$\rightarrow e  \rightarrow*$';
            \EndFor
                \State $events$=$seq.extractEvents()$;
                \State $last\_event$=$events[-1]$;
                \State $max\_delay$ = $max(\mathcal{D}_{delay}[`last\_event \rightarrow \mathcal{E}'])$ where $\mathcal{E}$ belongs to set of all possible events; 
                \If{$(current.time()$-$last\_event.time())>max\_delay$}
                    $seq$+=`$\rightarrow TO_i \rightarrow *$' where $i=1,2,...$;
                \EndIf
                \State $mismatch$=0;
                \If{$len(events)>n$}
                    \State $l$=$seq.extractLookaheadPairs()$;
                    \For{$pair$ in $l$}
                        \If{$pair$ not in $\mathcal{D}_{lookahead}$}
                            $mismatch$+=1;
                        \EndIf
                    \EndFor
                    \State $mismatch$=$mismatch/(n*(len(events)-(n+1)/2))$;
                    \If{$mismatch>\mathscr{t}$}
                        $seq$ is anomalous;
                    \ElsIf{$mismatch<\mathscr{t}$ and $events[-1]$==`$End$'}
                        \State $seq$ is normal;
                    \EndIf
                \EndIf
        \EndWhile
    \end{algorithmic}
\end{algorithm}

Once the length of $seq$ exceeds the window size $n$, we start extracting lookahead pairs from $seq$. Subsequently, we test for the presence or absence of the extracted lookahead pairs in $\mathcal{D}_{lookahead}$ to assign the mismatch score to $seq$.

\subsubsection{Assigning Mismatch Scores}
\label{subsec:assignment_scores}

Consider an example where the database contains the entries as shown in Table \ref{table:lookahead_all}, and we generate the following event sequence from an HTTP/2 request at a particular time instance $t_i$:

\begin{quote}
    $Start \rightarrow * \rightarrow Max\_Con\_Strm!0 \rightarrow * \rightarrow Ini\_Win\_Size!0 \rightarrow * \rightarrow win\_size\_incr!0 \rightarrow * \rightarrow Pref$
\end{quote}

For $n$=3, the lookahead pairs extracted from this event sequence are shown in Table \ref{table:lookahead_anomalous}.
\begin{table}
\renewcommand{\arraystretch}{1.3}
\caption{Lookahead pairs extracted from the test sequence}
\label{table:lookahead_anomalous}
\centering
\begin{tabular}{|p{1.75cm}|p{1.75cm}|p{1.75cm}|p{1.75cm}|}
\hline
& Index-1 & -2 & -3 \\ \hline
$Start$ & $*$ & $Max\_Con\_$ $Strm!0$ & \textcolor{red}{$*$} \\ 
\hline
$*$ & $Max\_Con\_$ $Strm!0$ \textbf{;} \textcolor{red}{$Ini\_Win\_Size$ $!0$} \textbf{;}  $win\_size\_incr$ $!0$  \textbf{;} $Pref$ & $*$ & \textcolor{red}{$Ini\_Win\_Size$ $!0$} \textbf{;} \textcolor{red}{$win\_size\_incr$ $!0$} \textbf{;} $Pref$\\ 
\hline
$Max\_Con\_$ $Strm!0$ & \textcolor{red}{$*$} & \textcolor{red}{$Ini\_Win\_Size$ $!0$} & \textcolor{red}{$*$} \\ 
\hline
$Ini\_Win\_Size$ $!0$ & $*$ & $win\_size\_incr$ $!0$ & $*$ \\ 
\hline
$win\_size\_incr$ $!0$ & $*$ & $Pref$ & \\ 
\hline
\end{tabular}
\end{table}
Since the lookahead pairs shown in red in Table \ref{table:lookahead_anomalous} do not exist in Table \ref{table:lookahead_all}, these are counted as mismatches. Subsequently, the mismatch score of the event sequence is computed as the ratio of the mismatch counts to the total possible mismatch counts. For instance, if we consider an event sequence of length $\mathcal{L}$ and a lookahead (window size) of $n$, the maximum pairwise mismatch count is given by
\begin{quote}
    $n(\mathcal{L}-n)+(n-1)+(n-2)+...+1=n(\mathcal{L}-(n+1)/2)$
\end{quote}
Thus, in the example discussed earlier, the mismatch score of the sequence is $7/21 (=0.33)$. The mismatch score of the event sequence is computed after every new event appended to the event sequence. It will be continued unless either of the following two conditions is met:
\\
\textbf{1.} The mismatch score of the sequence exceeds the predefined threshold. In this case, the corresponding HTTP/2 request from which the sequence is being generated will be treated as attack request.\\
\textbf{2.} The $End$ event is appended to the event sequence. After appending the $End$ event, if the mismatch score of the sequence is less than the predefined threshold, the corresponding HTTP/2 request will be treated as benign. However, if the score exceeds the threshold, the request will be treated as attack request.

\section{Experiments}
\label{sec:exp_eval}

\subsection{Learning Phase}
\label{sec:exp_learning}

There are no publicly available dataset for clear-text HTTP/2 traffic (h2c) \cite{praseed2021fuzzy}. Thus, we created a testbed similar to the one shown in Figure \ref{fig:testbed_exp} to capture clear-text normal HTTP/2 traffic during the learning phase and subsequently used it to build the lookahead pairs, and maximum delay databases:
\begin{figure}
\centering
\includegraphics[width=3.5in]{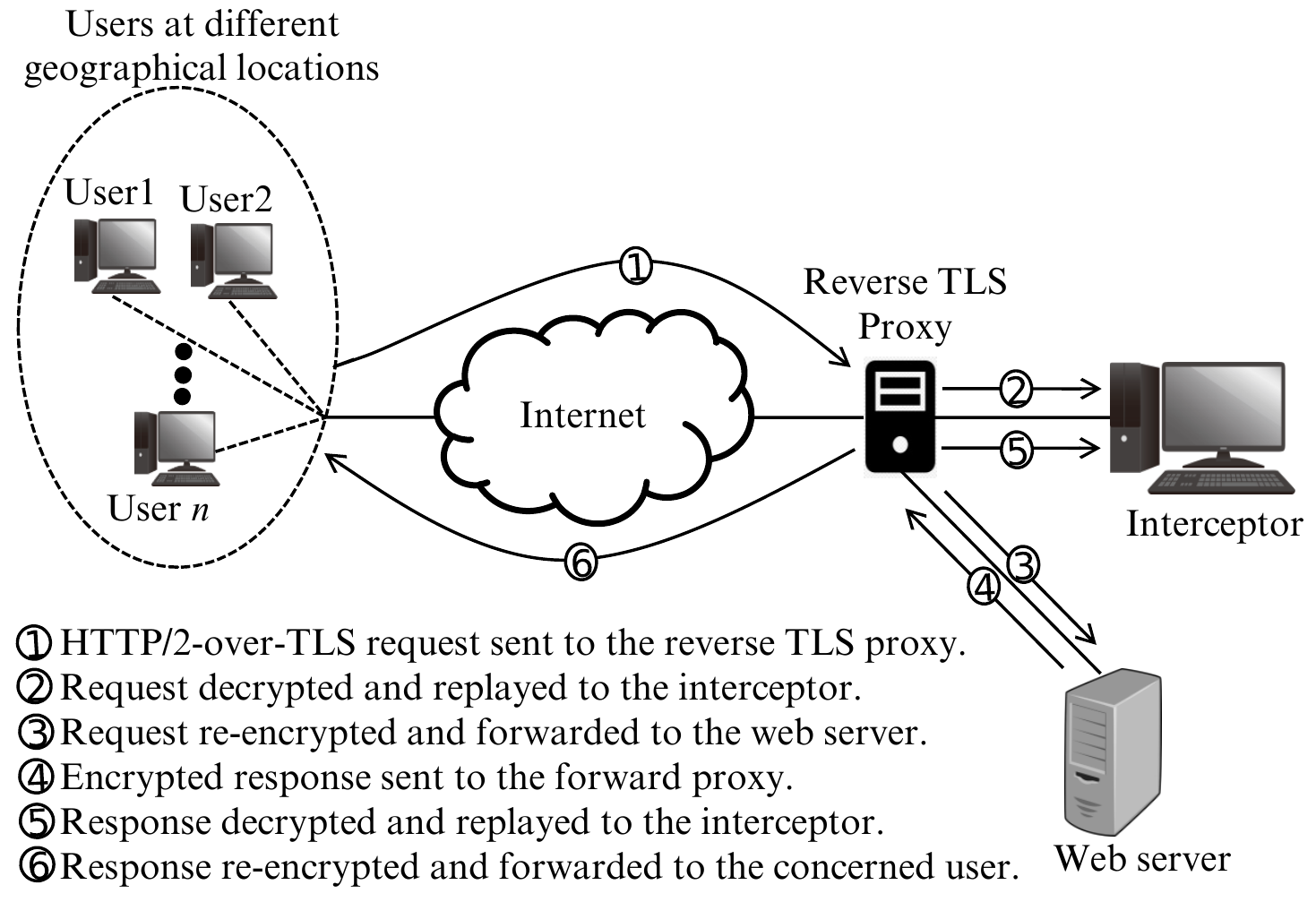}
\caption{Testbed setup for the experiments}
\label{fig:testbed_exp}
\end{figure}
\\
\textbf{Web Server:} The web server was running Ubuntu 20.04 LTS operating system and configured with Apache 2.4.51 to handle web requests. It had Intel Xeon E3 processor and 64 GB of physical memory. The web server was one of the academic servers of our institution, and it was hosting quiz assignments. The students were asked to attempt the assignments online for their academic grading. During the COVID-19 pandemic, the students visited the website from separate locations, and thus, the collected HTTP/2 traffic exhibited the property of real Internet behavior. We collected approximately 5 GB HTTP/2 traffic between October 8th, 2021, to October 14th, 2021.
\\
\textbf{User:} The users in our testbed were the students of our institute who accessed the web server to complete their assignments. They were asked to use the latest HTTP/2 supporting web browsers. We also practiced a few non-technical formalities to ensure that the students could not send any malicious traffic to poison the collected traffic.
\\
\textbf{Reverse TLS Proxy:} We configured a reverse TLS proxy application called PolarProxy\cite{netresec} in front of the web server. This proxy was owned by the website administration and was responsible for the tasks shown in Figure \ref{fig:testbed_exp}. 
%\textbf{First}, it decrypts the TLS connections originating from and destined to the web server\footnote{Outsourcing TLS traffic decryption is typically used by the website administrators to achieve better web server performance.} or the users. \textbf{Second}, it replays the decrypted HTTP/2 traffic to the interceptor on a designated port 57012 using \texttt{PCAP-OVER-IP} method. As a result, we obtain decrypted HTTP/2 traffic that was subsequently used for building the lookahead pair and the maximum delay database. \textbf{Third}, the proxy  re-encrypts the HTTP/2 traffic originating from and destined to the web server or the users.
\\
\textbf{Interceptor:} We designated a computer as the interceptor that intercepted the cleartext HTTP/2 traffic replayed by the proxy. The interceptor machine captured the inbound traffic to port 57012 using tcpdump. %and stored it in a \texttt{pcap} file.

\subsubsection{Capturing the normal behavior of HTTP/2}
\label{subsubsec:database_size}

\begin{figure}
\centering
\subfloat[Number of possible unique event sequences ]{\includegraphics[width=3.75in]{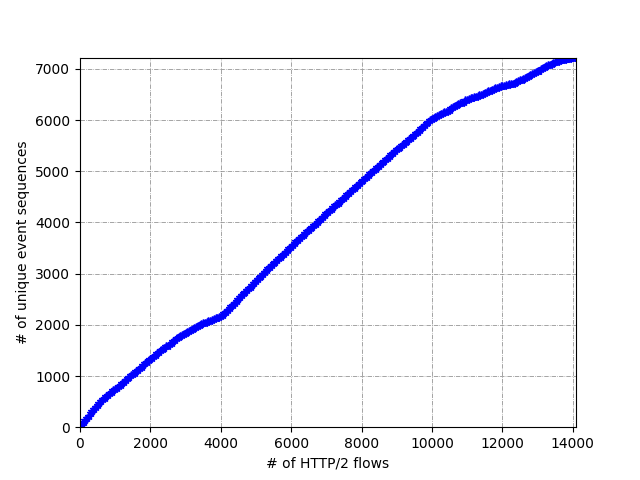}%
\label{fig:database_size_sequence}}
\hfil
\subfloat[Number of possible lookahead pairs for different values of $n$]{\includegraphics[width=3.75in]{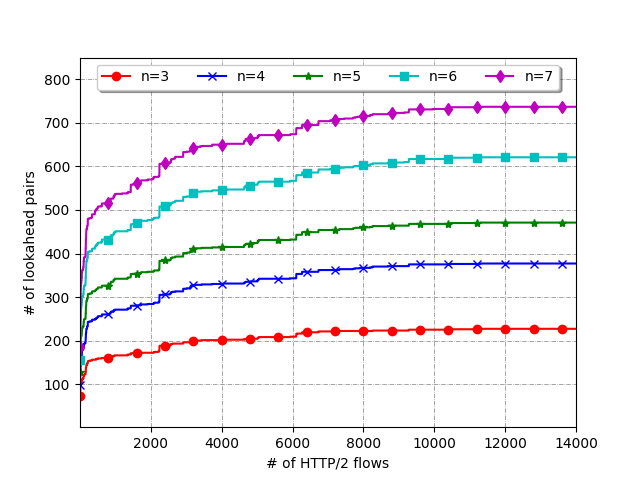}%
\label{fig:database_size_lookahead}}
\caption{Capturing normal HTTP/2 behavior. Please refer to the color version for better illustration.}
\label{fig:database_size}
\end{figure}

It is essential to decide what size of the database should be built to capture the normal behavior of HTTP/2. For this purpose, we collected 14094 HTTP/2 flows, and for each flow, we generated an event sequence. It resulted in 7221 unique event sequences. We also extracted the lookahead pairs from the unique event sequences and obtained 227, 377, 471, 621, and 737 unique lookahead pairs for window size ($n$)=3, 4, 5, 6, and 7, respectively. The plots in Figures \ref{fig:database_size_sequence} and \ref{fig:database_size_lookahead} show how the number of unique event sequences and lookahead pairs increases as we increase the number of HTTP/2 flows. We can notice from Figure \ref{fig:database_size_sequence} that the number of new event sequences kept increasing with the increase in the number of flows. Thus, it is not easy to estimate how much HTTP/2 traffic should be collected beforehand to obtain the unique event sequences possible in the case of HTTP/2 interactions. 

On the contrary, as can be noticed from Figure \ref{fig:database_size_lookahead}, virtually no new lookahead pairs were witnessed beyond a point $\mathscr{p}$ (approximately 9400 flows). Thus, we could assume that the database of lookahead pairs captured the virtually complete normal behavior of HTTP2 at $\mathscr{p}$. From this study, we can also establish that a database of lookahead pairs could capture the normal behavior of HTTP/2 with more completeness than a database of unique event sequences. %The size of the normal database gives an estimate of how much variability is possible in the normal behavior of HTTP/2.

\subsection{Detection Phase}

During the detection phase, we extended the setup shown in Figure \ref{fig:testbed_exp} by including one more entity and designating it as the attacker. The attacker machine was connected to the Internet and was responsible for sending the web server a small number of attack (Attack-1 to 5) requests at regular intervals. Moreover, the attacker terminated a connection\footnote{A connection was terminated using FIN-ACK TCP packet.} from which an attack request was sent as soon as 100 seconds progressed after the connection establishment. This was to ensure that such attack requests did not deplete the server's connection queue space. Closing these connections deliberately after 100 seconds did not alter the event sequences generated from the attack requests because our detection scheme could classify such requests as attack or benign way before the aforementioned 100 seconds (see Section \ref{subsubsec:detection_latency}). The benign requests were also generated during the detection phase while the students attempted their assignments.
%Using the modified testbed, we monitor the HTTP/2 traffic replayed by the proxy to the interpreter and classified them as benign or anomalous. 

The reverse TLS proxy in our testbed captured the benign and attack HTTP/2 requests generated by the students and the attacker, respectively, and forwarded their clear-text versions to the interceptor. Our detection scheme running on the interceptor then generated event sequences from the requests and subsequently classified them as normal or anomalous depending on their mismatch scores. We conducted this experiment into five different time intervals such that the window size $n$ was fixed, but the mismatch score threshold $\mathscr{t}$ was varied across the intervals. In this way, we could analyze the effect of varying $\mathscr{t}$ on the detection performance of the scheme. We used four metrics to evaluate the detection performance - Accuracy, False Positive Rate (FPR), Recall, and Precision.
% which are given by the Equations \ref{eq:acc}, \ref{eq:fpr}, \ref{eq:recall}, and \ref{eq:precision} respectively.
% \begin{equation}
%     \label{eq:acc}
%     Accuracy=\frac{TP+TN}{TP+TN+FP+FN}
% \end{equation}
% \begin{equation}
%     \label{eq:fpr}
%     FPR=\frac{FP}{TN+FP}
% \end{equation}
% \begin{equation}
%     \label{eq:recall}
%     Recall=\frac{TP}{TP+FN}
% \end{equation}
% \begin{equation}
%     \label{eq:precision}
%     Precision=\frac{TP}{TP+FP}
% \end{equation}
% In these equations, $TP$, $TN$, $FP$ and, $FN$ refer to true positive, true negative, false positive, and false negative respectively. $TP$ and $TN$ are the cases when a sequence is correctly classified. $FP$ is the case when a normal sequence is incorrectly classified as an anomalous sequence, and $FN$ is the case when an anomalous sequence is incorrectly classified as a normal sequence. 
The detection performance of the scheme in terms of these four metrics is shown in Table \ref{table:detect_perform}. We can notice from the table that $Accuracy$ and $Precision$ slightly increased as we increased $\mathscr{t}$ to 0.02 from 0.01. It was because of the reduced number of $FP$ cases (a normal sequence incorrectly classified as an anomalous sequence) and the increased number of $TN$ cases. Due to the same reason, $FPR$ decreased as we increased $\mathscr{t}$ to 0.02 from 0.01. However, as we further increased $\mathscr{t}$, $Accuracy$ decreased because of more $FN$ cases (an anomalous sequence incorrectly classified as a normal sequence). Moreover, $Recall$ and $Accuracy$ decreased drastically as we increased $\mathscr{t}$ to 0.03 from 0.02. It was because the anomalous sequences with a mismatch score in the range [0.02, 0.03] were considered as normal resulting in a large number of $FN$ cases. Due to these misclassifications, we obtained lower detection accuracy. The same holds true as we further increased the value of $\mathscr{t}$.

\begin{table*}
\renewcommand{\arraystretch}{1.3}
\caption{Detection performance}
\label{table:detect_perform}
\centering
%\begin{tabular}{|p{0.8cm}|c|c|p{0.3cm}|p{1cm}|p{0.3cm}|p{0.6cm}|p{1cm}|}
\begin{tabular}{|c|c|c|c|c|c|c|c|c|}
\hline
\multirow{2}{*}{Interval} & \multicolumn{2}{c|}{\#\ of Flows} & \multirow{2}{*}{Mismatch score threshold $\mathscr{t}$} & \multirow{2}{*}{Window size $n$} &\multirow{2}{*}{Accuracy} & \multirow{2}{*}{FPR} & \multirow{2}{*}{Recall} & \multirow{2}{*}{Precision} \\ \cline{2-3}
& Benign & Attack & & & & & & \\ \hline
1$^{st}$ & 12423 & 3242 & 0.01 & 5 & 98.55 & 1.83 & 100.00 & 93.46 \\ \hline
2$^{nd}$ & 12423 & 3225 & 0.02 & 5 & 98.65 & 1.71 & 100.00 & 93.83 \\ \hline
3$^{rd}$ & 12120 & 3222 & 0.03 & 5 & 93.39 &	1.64 &	074.71 & 92.36 \\ \hline
4$^{th}$ & 10605 & 2860 & 0.04 & 5 & 88.91 & 1.64 & 053.88 & 89.85 \\ \hline
5$^{th}$ & 10620 & 3070 & 0.05 & 5 & 82.80 & 1.62 & 006.08 & 43.23 \\ \hline
\end{tabular}
\end{table*}

\subsubsection{Mismatch scores of normal and anomalous event sequences}

\begin{figure}
\centering
\includegraphics[width=3.75in]{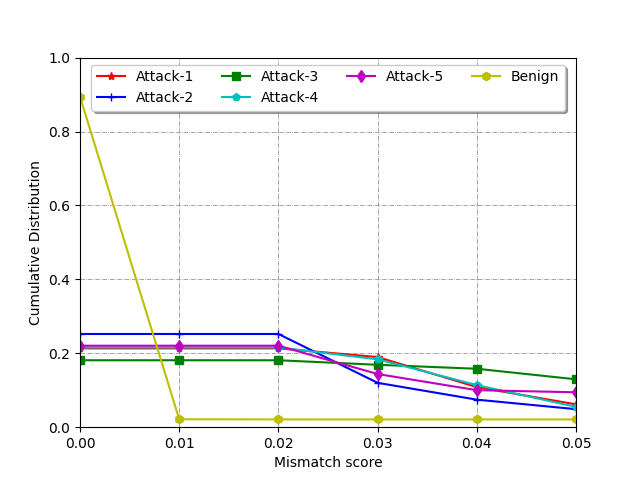}
\caption{CDF plot of the mismatch scores of normal and anomalous sequences for $n$=5. Please refer to the color version for better illustration.}
\label{fig:mismatch_score_normal_anomalous}
\end{figure}

It can be noticed from Figure \ref{fig:mismatch_score_normal_anomalous} that the number of anomalous sequences with a mismatch score in the range [0, 0.02] is almost the same across this range, and the mismatch scores of virtually all normal sequences laid in the range [0, 0.01]. Thus, to minimize the number of $FP$ without any significant increase in $FN$, the mismatch score threshold $\mathscr{t}$ should be chosen in the range [0.01, 0.02]. As we increase $\mathscr{t}$ beyond 0.02, the number of $FN$ would start increasing significantly without any substantial reduction in the number of $FP$ due to which the detection performance would degrade. Thus, choosing $\mathscr{t}$ in the range [0.01, 0.02] would result in the best detection performance of the scheme, as also apparent from Table \ref{table:detect_perform}.

\subsubsection{Detection performance v/s different values of $n$}

\begin{figure}
\centering
\includegraphics[width=3.75in]{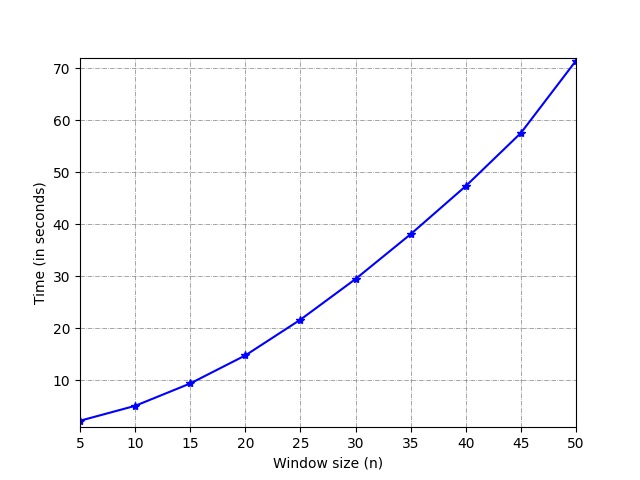}
\caption{Time required to extract lookahead pairs for different $n$}
\label{fig:lookahead_pairs_time}
\end{figure}

To analyze the effect of varying the window size $n$ on the detection performance of the scheme, we experimented with four more time intervals of 8 hours each such that the mismatch score threshold $\mathscr{t}$ was fixed, but $n$ was varied across the intervals. We show the results of this experiment in Table \ref{table:detect_perform_vary_n}. 
\begin{table}
\renewcommand{\arraystretch}{1.3}
\caption{Effect of varying $n$ on detection performance}
\label{table:detect_perform_vary_n}
\centering
\begin{tabular}{|c|c|c|c|c|c|c|c|}
\hline
Window size ($n$) & Accuracy & FPR & Recall & Precision \\ \hline
3 & 98.55 &	1.79 &	100 & 92.94 \\ \hline
5 & 98.35 &	2.04 &	100 & 92.03 \\ \hline
7 & 98.68 &	1.63 &	100 & 93.52 \\ \hline
9 & 98.89 & 1.37 & 100 & 94.52 \\ \hline
\end{tabular}
\end{table}
We can notice from the table that varying the window size does not virtually affect the proposed scheme's detection accuracy. However, we also observed from our experiments that the longer the chosen window size, the longer time was required to extract the lookahead pairs from an event sequence, as shown in Figure \ref{fig:lookahead_pairs_time}. 
This results in the delayed classification of an event sequence as a normal or anomalous sequence. Thus, choosing a smaller window size is essential so that the mismatch scores can be calculated and the corresponding decision can be made as soon as possible.

\subsubsection{Detection latency and computational overhead}
\label{subsubsec:detection_latency}

\begin{figure}
\centering
\includegraphics[width=3.75in]{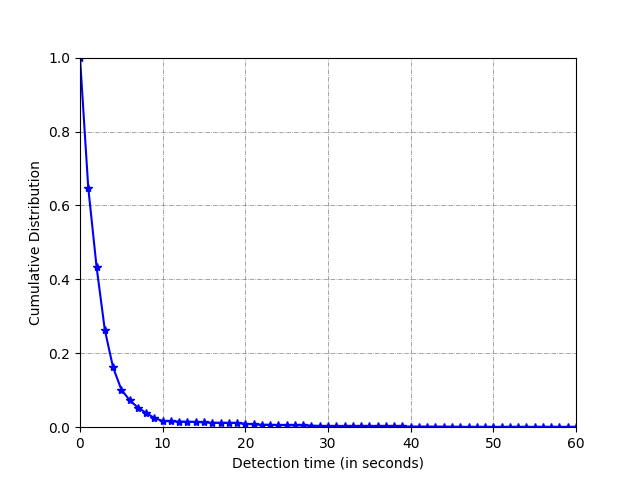}
\caption{CDF plot of the time required to detect anomalous event sequences}
\label{fig:detection_latency}
\end{figure}

It is essential for an active detection scheme to detect anomalies as soon as possible so that an alarm can be raised and necessary preventive measures can be taken. Thus, we also measured how fast our proposed detection scheme could detect anomalous event sequences. Figure \ref{fig:detection_latency} shows the result of this measurement. We can notice from the figure that the detection scheme could detect all anomalous sequences in virtually less than 30 seconds. 

\begin{figure}
\centering
\includegraphics[width=3.75in]{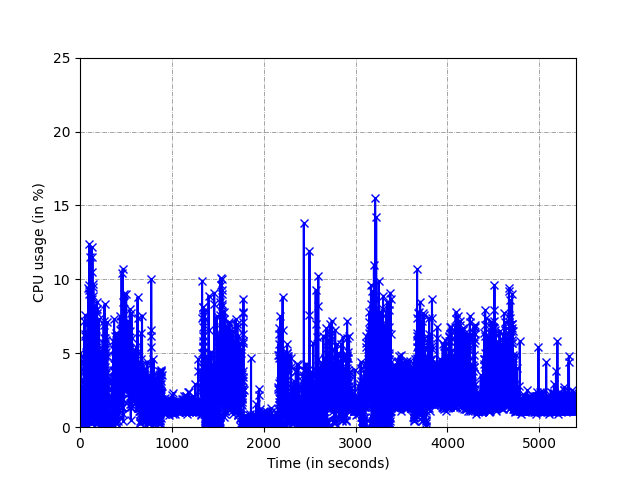}
\caption{Computational overhead}
\label{fig:cpu_usage}
\end{figure}

While the detection scheme was running at the Interpreter, we also monitored for 1.5 hours the computational overhead incurred due to its execution.
In particular, we logged the computer's overall CPU\footnote{The Interpreter computer had AMD Ryzen 7 5800H processor.} utilization at regular intervals of 1 second using \texttt{psutil} and plotted the graph shown in Figure \ref{fig:cpu_usage}. We can notice from the figure that the CPU usage, while the detection scheme was running, was less than $\approx$7\% (dense blue area) in most of the time intervals, except for some intervals wherein the CPU usage reached up to $\approx$16\%. Thus, due to the small computational overhead of the proposed detection scheme, it can be an ideal solution for detecting Slow HTTP/2 DoS attacks in real-time.

\section{Conclusion and Future Work}
\label{sec:conclusion}

Slow Rate DoS attacks are a matter of grave concern for server administrators because of two prominent reasons. First, these attacks require significantly less computational power and thus, can be launched even from mobile devices. Second, these attacks generate minimal traffic, due to which they are highly stealthy. The effect of these attacks on HTTP/2 servers on the Internet had not been analyzed earlier. Moreover, the known mechanisms to counter the attacks against HTTP/1.1 can not counter them against HTTP/2. Thus, in this work, we attempted to bridge these gaps by first performing an empirical evaluation of Slow HTTP/2 DoS attacks on the Internet and subsequently proposing a scheme to detect these attacks in real-time. Our experiments showed that several HTTP/2 servers on the Internet \textbf{delay} closing the connections from which incomplete requests are sent, thus being vulnerable to the Slow Rate DoS attacks.
%Our proposed detection scheme considers an HTTP/2 interaction as a sequence of events and checks for any anomalies in this sequence to detect the anomalous HTTP//2 requests. 
Our experiments to test the performance of the proposed detection scheme showed that it could accurately detect the attacks in real-time with a marginal computational overhead. In the future, we plan to work on the vulnerability assessment of HTTP/3, the upcoming version of HTTP. As of January 2022, HTTP/3 has not been standardized, and it currently holds the status of an Internet Draft.

\bibliographystyle{IEEEtran}
\bibliography{bare_jrnl_compsoc}

% \appendices
% \section{XYZ}

% \section{}

\ifCLASSOPTIONcompsoc
  \section*{Acknowledgments}
\else
  \section*{Acknowledgment}
\fi

The authors would like to thank the Netresec team for supporting this research with a free L1 license of PolarProxy.

\ifCLASSOPTIONcaptionsoff
  \newpage
\fi

\begin{IEEEbiography}[{\includegraphics[width=1in,height=1.25in,clip,keepaspectratio]{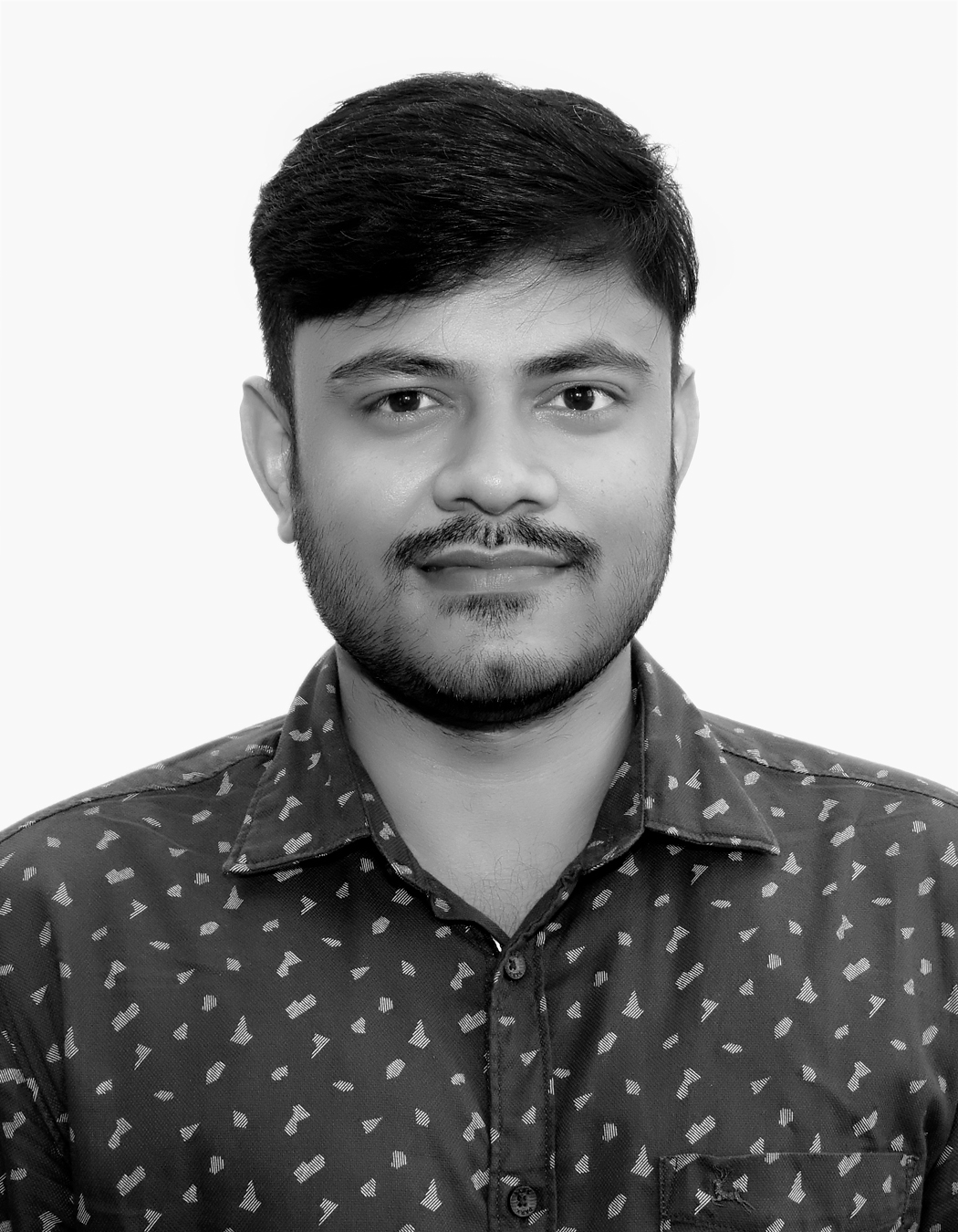}}]{Nikhil Tripathi} is currently working as an Assistant Professor in Computer Science \& Engineering Group at the Indian Institute of Information Technology, Sri City, India. Prior to his current role, he worked as a cyber security researcher at Fraunhofer Institute for Secure Information Technology (SIT), Germany. He earned his Ph.D. in Computer Science and Engineering from the Indian Institute of Technology Indore, India, in 2019. He primarily works on the vulnerability assessment of popular application layer protocols to explore possible DoS vulnerabilities and devise novel defense mechanisms to counter large-scale DoS attacks. He has published research papers in various reputed journals such as ACM Computing Surveys, Computers \& Security, and conferences like INFOCOM, ARES, and COMSNETS. He also holds professional memberships of societies such as IEEE, IEEE Communications Society, and ACM.
\end{IEEEbiography}

\end{document}